\documentclass[journal]{IEEEtran}
\usepackage{amsmath,amssymb,amsfonts}
\usepackage{physics}
\usepackage{graphicx}
\usepackage{booktabs}
\usepackage{algorithm}
\usepackage{algpseudocode}
\usepackage{cite}
\usepackage{url}
\usepackage{xcolor}
\usepackage{quantikz}
\usepackage{orcidlink}
\usepackage{hyperref}
\usepackage{tikz}
\usetikzlibrary{arrows.meta, positioning, backgrounds, calc, fit, shapes.geometric}

\definecolor{BoxBlue}{HTML}{80e7fe}
\definecolor{DarkBlue}{HTML}{166083}
\definecolor{BoxPurple}{HTML}{662f89}
\definecolor{BoxMaroon}{HTML}{f88c70}
\definecolor{CircleYellow}{HTML}{ffc000}
\IEEEoverridecommandlockouts

\title{Quantum computing-based solver for interacting power grids}

\author{

    \IEEEauthorblockA{Ankit Kumar Das\orcidlink{0009-0001-2252-2657}$^{1}$, Durgesh Pandey\orcidlink{0009-0001-0877-1980}$^{1}$, and P. Arumugam\orcidlink{0000-0001-9624-8024}$^{1,2}$}
        
    \IEEEauthorblockA{$^1$Department of Physics, Indian Institute of Technology Roorkee, Roorkee, Uttarakhand 247667, India}
    
    \IEEEauthorblockA{$^2$Centre for Photonics and Quantum Communication Technology, Indian Institute of Technology Roorkee, Roorkee 247667, India}
    
    \IEEEauthorblockA{Email: ankitk\_das@ph.iitr.ac.in, durgesh\_p@ph.iitr.ac.in,  arumugam@ph.iitr.ac.in}
}

\begin{document}
\maketitle

\begin{abstract}
The proliferation of power electronics in multi-terminal transmission grids has increasingly led to harmonic distortions and dynamic instabilities. While Resonance Mode Analysis (RMA) provides deep insights into these system resonances, evaluating the critical modes of large-scale grids presents a severe computational bottleneck. Classical iterative techniques must continuously diagonalize massively high-dimensional, non-Hermitian admittance matrices across a wide frequency spectrum, a process that rapidly exhausts classical memory and processing limits. To overcome this scaling barrier, we propose a novel quantum-classical hybrid methodology that natively maps the transmission grid's admittance matrix onto a Quantum Processing Unit (QPU). Because the grid's matrix is non-Hermitian, standard quantum eigensolvers are insufficient; thus, we employ the Real Variance-based Variational Quantum Eigensolver (RVVQE) algorithm to accurately extract the complex eigenvalues that represent the system's modes. Validated against a standard 5-bus transmission system, the quantum-derived critical-resonance modal impedances demonstrate near-perfect alignment with the exact classical frequency responses. Crucially, by encoding the grid's state logarithmically into quantum memory, this methodology bypasses classical RAM limitations. The successful implementation of the RVVQE framework not only bridges the mathematical topologies of dissipative electrical grids and open quantum systems but also provides a profoundly scalable architecture capable of diagnosing resonance instabilities in massive, continental-scale networks that currently exceed classical computational boundaries.
\end{abstract}

\begin{figure*}[htbp]
    \centering
    \includegraphics[width=0.9\linewidth]{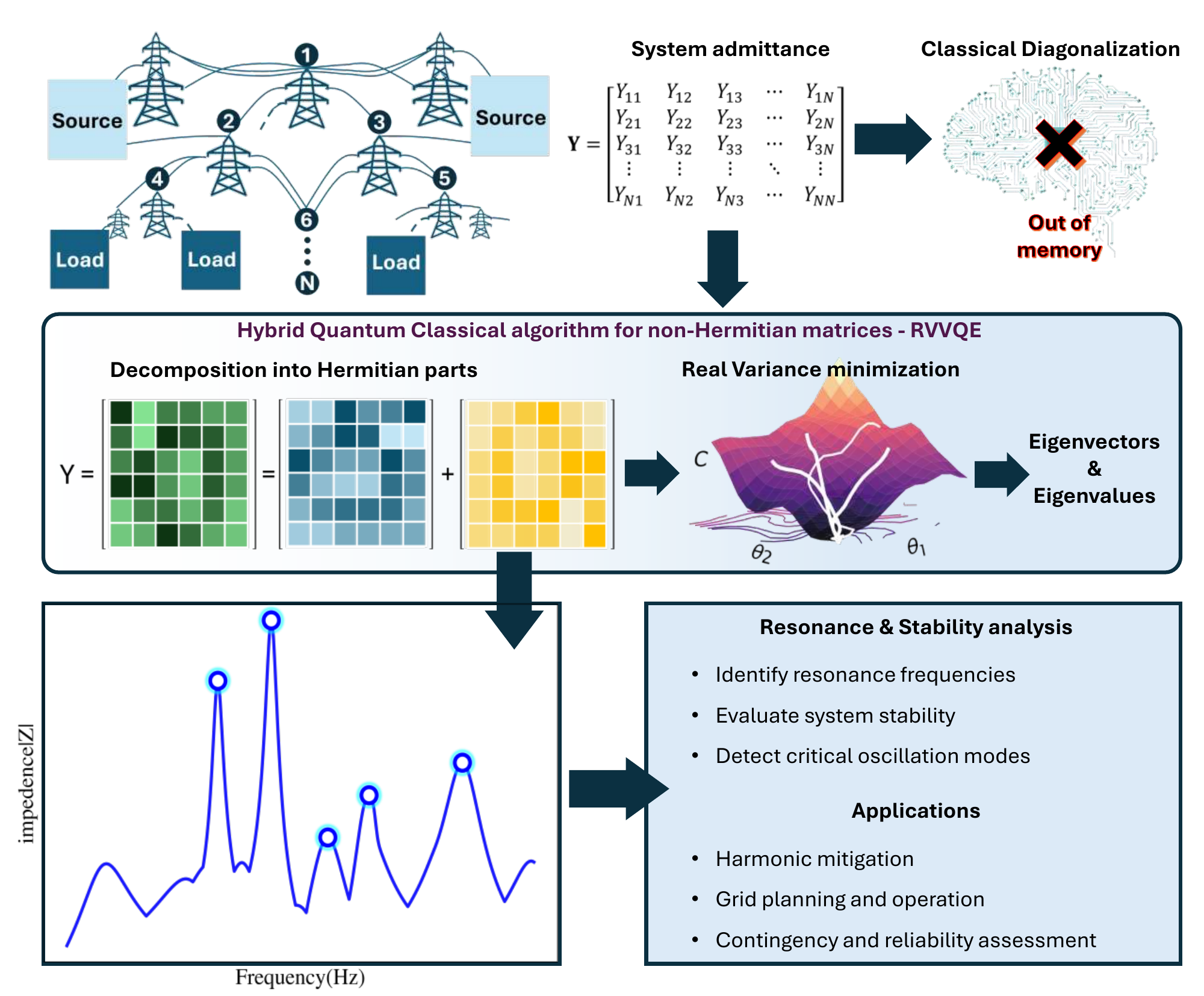}
        \caption{Overview of the proposed hybrid quantum-classical framework for resonance mode analysis (RMA) of power transmission grids. The system admittance matrix is decomposed into Hermitian components and solved using the Real Variance-based Variational Quantum Eigensolver (RVVQE) to obtain the eigenvalues and eigenvectors required for resonance and stability analysis.}
    \label{fig:graph_abstract}
\end{figure*}



\section{Introduction}

The increasing use of renewable energy sources and high-voltage power electronic converters has made modern transmission grids more complex and dynamic \cite{sun2017renewable}. Although these technologies improve power transfer and system flexibility, they also introduce high-frequency harmonic currents. These harmonics interact with the passive components of the network and can lead to resonance, increased harmonic distortion, and system instability \cite{cui2012modal}. Frequency-scan analysis based on the voltage-node method is commonly used to identify resonance frequencies~\cite{CARTIEL2026486}. However, as the size of the transmission network increases, this method alone cannot provide enough information about where the resonance occurs or how it affects different buses in the system \cite{xu2005harmonic}.

To overcome this limitation, Resonance Mode Analysis (RMA) was introduced. RMA performs eigenvalue decomposition of the system admittance matrix over a range of frequencies to identify the dominant resonance modes that influence system stability \cite{zhan2019frequency,xu2005harmonic,huang2007application}. The method has been successfully applied to several benchmark power systems, including the IEEE 14-bus, 24-bus, 118-bus, and 300-bus test networks \cite{ieee14bus,ieee24bus,ieee118bus,ieee300bus}. From the eigenvalues and eigenvectors, participation factors can be calculated to locate resonance sources and study the stability of converter-dominated power systems \cite{orellana2021stability,Zhou2025trps,ebrahimzadeh2018bus}.

A major challenge in RMA is the computation of the eigenvalues of the admittance matrix. Although the matrix is symmetric, it is generally non-Hermitian because it contains both active and reactive power components. For large transmission networks with thousands of buses, storing and diagonalizing these matrices becomes computationally expensive. A classical computer requires $\mathcal{O}(n^2)$ memory to store an $n \times n$ matrix, while direct eigenvalue decomposition scales as $\mathcal{O}(n^3)$ \cite{saad2011numerical,golub1996matrix}. Iterative methods, such as the non-Hermitian Lanczos algorithm, reduce the computational cost for sparse matrices but still require storing large Krylov subspace vectors in memory \cite{cartiel2023faster,day1997efficient}. As the network size grows to tens of thousands of buses \cite{activsg70k}, these memory requirements become a major limitation.

Quantum computing provides a different way to solve large computational problems. Since Feynman's proposal in 1982, quantum computers have been considered a promising platform for simulating complex physical systems more efficiently than classical computers \cite{feynman1982simulating}. Instead of storing the system state in a large number of classical memory locations, quantum computers encode the information into the amplitudes of qubits. An $n$-dimensional system can be represented using only $N=\lceil\log_2 n\rceil$ qubits \cite{giovannetti2008quantum}. For example, a transmission grid with about 70,000 buses can be represented using only 17 qubits because $2^{17}=131\,072$. This logarithmic scaling greatly reduces the memory required to represent large systems.

However, most existing quantum eigensolver algorithms, such as the Variational Quantum Eigensolver (VQE), are designed only for Hermitian matrices \cite{peruzzo2014variational,pooja_gc_2021}. Therefore, they cannot be directly applied to the non-Hermitian admittance matrices used in resonance mode analysis. In this work, we use the Real Variance-based Variational Quantum Eigensolver (RVVQE) to overcome this limitation \cite{pandey2026rvvqe}. The non-Hermitian admittance matrix is first decomposed into two Hermitian matrices. These matrices are then mapped onto a quantum circuit, and a classical optimizer is used to minimize a real-valued variance function to obtain the eigenvalues and eigenstates of the original non-Hermitian system \cite{pandey2026rvvqe,ashutosh_complex_2025}. Using this approach, we show that quantum computing can accurately calculate the resonance modes of power systems while avoiding the large memory requirements of classical methods. This provides a scalable framework for resonance analysis of future large-scale transmission networks.
\section{Theoretical Formalism}
\begin{figure*}[htbp]
\centering
\begin{tikzpicture}[
    node distance=1.2cm and 1.0cm,
    stdBlock/.style={
        rectangle, 
        draw=DarkBlue, 
        fill=white, 
        text=black, 
        line width=1.5pt, 
        align=center, 
        font=\normalsize, 
        minimum height=1.4cm, 
        minimum width=4.2cm, 
        inner sep=6pt
    },
    qpuBlock/.style={
        rectangle, 
        draw=BoxPurple, 
        fill=white, 
        text=black, 
        line width=1.5pt, 
        align=center, 
        font=\normalsize, 
        minimum height=1.4cm, 
        minimum width=4.2cm, 
        inner sep=6pt
    },
    decision/.style={
        diamond,
        draw=BoxPurple,
        fill=white,
        text=black,
        line width=1.5pt,
        align=center,
        font=\normalsize,
        inner sep=2pt,
        aspect=1.8
    },
    arrow/.style={-{Stealth[scale=1.2]}, draw=black, line width=2pt},
    thickArrow/.style={-{Stealth[scale=1.3]}, draw=black, line width=2.5pt}
]

    \node[circle, draw=DarkBlue, fill=white, text=black, line width=1.5pt, font=\normalsize, align=center, inner sep=6pt, minimum size=3.0cm] 
        (Start) at (0, 4.0) {Frequency\\Sweep\\$f^{(0)} \to f^{(\mathrm{mx})}$};
        
    \node[stdBlock, right=of Start, xshift=0.4cm] (Y) {
        \textbf{System Decomposition} \\ 
        $Y_{B,f} = H_f + iK_f$ \\
        \small Map to Pauli Strings
    };

    \node[qpuBlock, below=of Y, yshift=-1.4cm] (Init) {
        \textbf{State Initialization} \\
        Prepare $|\psi(\theta)\rangle$
    };

    \node[qpuBlock, below=of Init, yshift=-0.6cm] (Measure) {
        \textbf{QPU Measurement} \\
        Evaluate $\langle H_f^k \rangle, \langle K_f^k \rangle$
    };

    \node[decision, right=of Measure, xshift=1.8cm] (Verify) { 
        Is $C_{\mathrm{var}}(\theta) \approx 0$?
    };

   \node[draw=BoxPurple!30, fill=none, line width=1.5pt, rectangle, left=of Init, xshift=0.5cm, yshift = -1.5cm, minimum width=4.2cm, inner sep=4pt] (CalcPre) {
    \begin{tabular}{c}
        \textbf{ QUANTUM } \\
        \textbf{ HYBRID } \\ \\ 
        \begin{tabular}{|c|}
            \hline
            \\  \textbf{\quad CPU \quad} \\ \\ 
            \hline
        \end{tabular} \\
        \Huge{$\Downarrow$ $\Uparrow$}  \\
        \begin{tabular}{|c|}
            \hline
            \\  \textbf{\quad QPU \quad} \\ \\ 
            \hline
        \end{tabular}
    \end{tabular}
    };
    
    \node[qpuBlock, above=of Verify, yshift=0.3cm] (Map) {
        \textbf{Modal Mapping} \\
        Compute Eigenvalue $|\lambda|$ \\
        \& Eigenvector for $Z_{\mathrm{cm},f}$
    };

    \node[circle, above=of Map, , yshift=0.5cm, draw=CircleYellow, fill=white, text=black, line width=2.5pt, font=\normalsize, align=center, inner sep=6pt, minimum size=3.0cm] 
        (End) {Stability\\Assessment \\ \& Output \\ $Z_{\mathrm{cm},f}$};

    \begin{scope}[on background layer]
        \node[draw=BoxBlue, fill=BoxBlue!30, line width=2pt, rounded corners=4pt, inner sep=12pt, fit={(Start)(Y)}] (BoxFormulation) {};
        \node[text=DarkBlue, font=\bfseries\sffamily\normalsize, anchor=south west] at (BoxFormulation.north west) {Formulation};

        \node[draw=BoxPurple, fill=BoxPurple!30, line width=2pt, rounded corners=4pt, inner sep=12pt, fit={(Init) (CalcPre) (Measure) (Verify) (Map)}] (BoxRVVQE) {};
        \node[text=BoxPurple, font=\bfseries\sffamily\normalsize, anchor=south west] at (BoxRVVQE.north west) {RVVQE Loop};

        \node[draw=BoxMaroon, fill=BoxMaroon!30, line width=2pt, rounded corners=4pt, inner sep=12pt, fit={(End)}] (BoxOutput) {};
        \node[text=BoxMaroon, font=\bfseries\sffamily\normalsize, anchor=south west] at (BoxOutput.north west) {Output};
    \end{scope}

    \draw[arrow] (Start) -- (Y);
    
    \draw[thickArrow] (Y.south) -- (Init.north);
    
    \draw[arrow] (Init) -- (Measure);
    \draw[arrow] (Measure) -- (Verify);
    
   \draw[arrow, dashed, BoxPurple] (Verify.north west) -- node[midway, below left, font=\small\bfseries, text=BoxPurple, inner sep=2pt] {No (update $\theta$)} (Init.east);
    
    \draw[arrow] (Verify.north) -- node[midway, right, font=\small\bfseries, text=BoxPurple] {Yes} (Map.south);
    
    \draw[thickArrow] (Map.north) -- (End.south);


\end{tikzpicture}
\caption{Workflow of the proposed hybrid quantum-classical (HQC) framework for resonance mode analysis. The non-Hermitian admittance matrix is decomposed into Hermitian components and mapped to Pauli operators. The RVVQE algorithm iteratively prepares the quantum state, performs QPU measurements, and updates the circuit parameters until convergence, after which the resonance modes are extracted for stability assessment.}
\end{figure*}
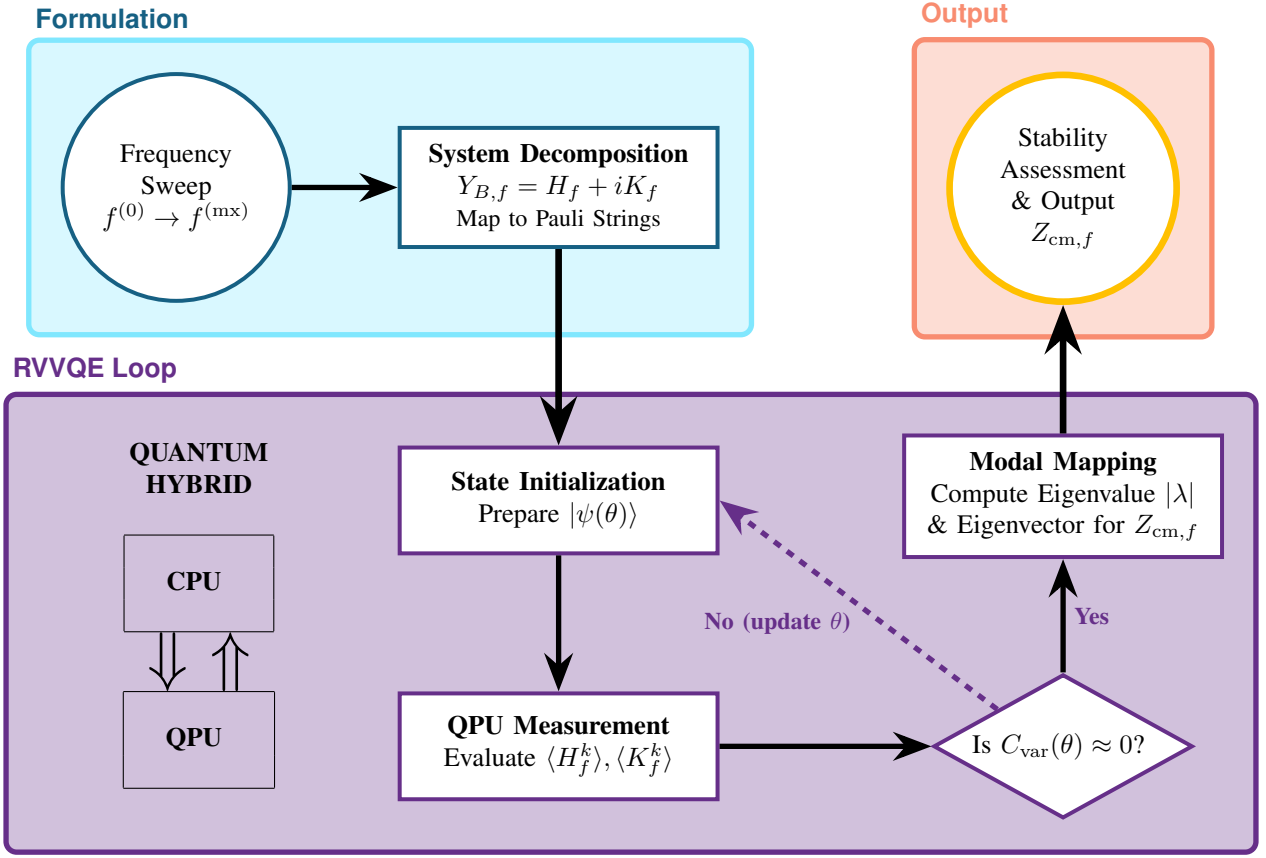
\subsection{Classical problem in transmission grid}

To evaluate resonances in transmission grids, engineers start with a frequency-scan analysis using the voltage-node method. For a given frequency $f$, the relationship between the grid's nodal voltages and injected currents is:
\begin{equation}
V_{B,f} = Y_{B,f}^{-1}I_{B,f} = Z_{B,f}I_{B,f}
\end{equation}
Here, $V_{B,f}$ is the voltage vector, $I_{B,f}$ is the current injection vector, $Y_{B,f}$ is the system admittance matrix, and $Z_{B,f} = Y_{B,f}^{-1}$ is the impedance matrix. By observing how the impedance matrix changes with frequency, we can identify grid resonances. These resonances manifest as peak impedance values that lead to high bus voltages \cite{xu2005harmonic}. Mathematically, these peaks occur when one or more eigenvalues of the admittance matrix $Y_{B,f}$ approach zero, making the matrix nearly singular and causing the corresponding modal impedances to become large.

To perform these calculations more efficiently for large grids, Resonance Mode Analysis (RMA) is used. It computes the eigenvalues and eigenvectors of the admittance matrix at each frequency step, allowing resonance conditions to be identified by tracking the eigenvalues across the frequency range.\cite{xu2005harmonic, cartiel2023faster}. The admittance matrix is decomposed as
\begin{equation}
Y_{B,f}=R_f\Lambda_{Y,f}L_f,
\end{equation}
which transforms the nodal equations into the modal domain:
\begin{equation}
U_f=\Lambda_{Y,f}^{-1}J_f.
\end{equation}
where $U_{f} = L_{f}V_{B,f}$ and $J_{f} = L_{f}I_{B,f}$ are the modal voltage and current vectors. The diagonal matrix $\Lambda_{Y,f}$ holds the eigenvalues of $Y_{B,f}$. The matrices $R_{f}$ and $L_{f}$ contain the right (in columns) and left (in rows) eigenvectors \cite{saad2011numerical, golub1996matrix}. Because the grid's admittance matrix is complex symmetric, these eigenvectors relate to each other as:
\begin{equation}
R_{f} = L_{f}^{-1} = L_{f}^{T}
\end{equation}

The eigenvalues of $Y_{B,f}$ represent all the possible modes of the grid at a certain frequency. Resonance modes are identified by tracking the admittance matrix's eigenvalues across the frequency range. A resonance occurs when an eigenvalue reaches a local minimum, corresponding to a maximum in the associated modal impedance \cite{li2018rapid,xu2005harmonic}. These minimums match the maximum values of the eigenvalues of the impedance matrix, which are called resonance modal impedance ($Z_{mj,f} = 1/\lambda_{Yj,f}$). The mode with the highest impedance at a given frequency is referred to as the critical mode. The resonance frequencies of the system are exactly at the frequency where these critical mode peaks. 

RMA also shows how individual buses interact with these critical resonance modes \cite{huang2007application}. This interaction is measured using participation factors (PFs), which are calculated from the right and left eigenvectors:
\begin{equation}
PF_{bj,f} = R_{bj,f}L_{jb,f}
\end{equation}

Beyond resonance identification, this modal approach can also be used to assess small-signal stability, particularly for converter-driven instabilities \cite{li2018probabilistic, yang2022siso}. According to the positive-mode-damping (PMD) stability criterion \cite{orellana2021stability}, the grid is stable if the product of the real part of the critical modal impedance and the frequency derivative of its imaginary part is negative. For the critical modal impedance $Z_{cm,f}=R_{cm,f}+iX_{cm,f}$, this condition is expressed as
\begin{equation}
\frac{\partial X_{cm,f}}{\partial f}\Bigg|_{f=f_r}R_{cm,f_r}<0.
\end{equation}

\subsection{Quantum Embedding and the RVVQE Algorithm}

To perform this calculation on a quantum computer, the classical admittance matrix is mapped into a quantum framework. Because standard quantum algorithms, including the Variational Quantum Eigensolver (VQE), are designed for Hermitian operators, they are ineffective in recovering correct eigenvalues for non-Hermitian matrices \cite{pandey2026rvvqe}. To solve this, we apply a systematic formulation based on a Real Variance-based Variational Quantum Eigensolver (RVVQE) algorithm, which is specifically designed for non-Hermitian operators \cite{pandey2026rvvqe}.

The computational process begins by extracting the $n$-bus transmission grid data and defining the frequency spectrum of interest. At each discrete frequency step $f$, the non-Hermitian admittance matrix $Y_{B,f}$ is calculated using the established white-box and black-box models of the grid components \cite{cartiel2023faster}. This matrix is then mathematically decomposed into two purely Hermitian matrices, $H_f$ and $K_f$, such that $Y_{B,f}=H_f+iK_f$. These Hermitian components are mapped into Pauli strings for measurement on an $N$-qubit quantum hardware system, where $N = \lceil \log_2 n \rceil$. 
\begin{figure}
    \centering
    \includegraphics[width=1\linewidth]{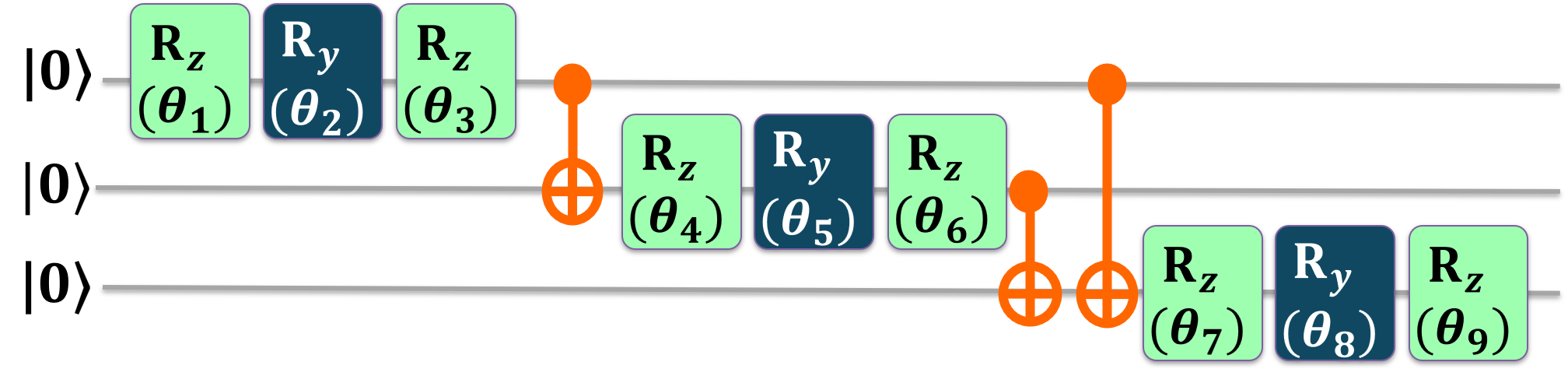}
    \caption{Three-qubit unitary rotation entangled ansatz with nearest-neighbor CNOT gates. The complete rotation for each qubit allows exploration of full Hilbert space spanned by 3 qubits.}
    \label{fig:ansatz_3q_ent}
\end{figure}
Next, a parameterized quantum circuit, known as the ansatz, prepares a trial state $|\psi(\theta)\rangle$.  Instead of minimizing standard energy, the RVVQE algorithm calculates a real-valued variance cost function to identify the critical eigenstates. This cost function is defined as the sum of the variances of the two Hermitian components:
\begin{equation}
C_{var}(\theta) = \langle H_f^2 \rangle_{\theta} - \langle H_f \rangle_{\theta}^2 + \langle K_f^2 \rangle_{\theta} - \langle K_f \rangle_{\theta}^2.
\end{equation}
A classical optimizer continuously updates the parameters $\theta$ via a hybrid quantum-classical workflow to minimize this cost function \cite{pandey2026rvvqe}. Once the cost function converges to zero (within an acceptable tolerance $\epsilon$), the optimal parameters $\theta^*$ are identified. The critical complex eigenvalue $\lambda_{n,f}$ is calculated by adding the individual expectation values, yielding the resonance modal impedance $Z_{cm,f} = 1/\lambda_{n,f}$. 

\begin{figure}[htbp]
    \centering
    \includegraphics[width=\columnwidth]{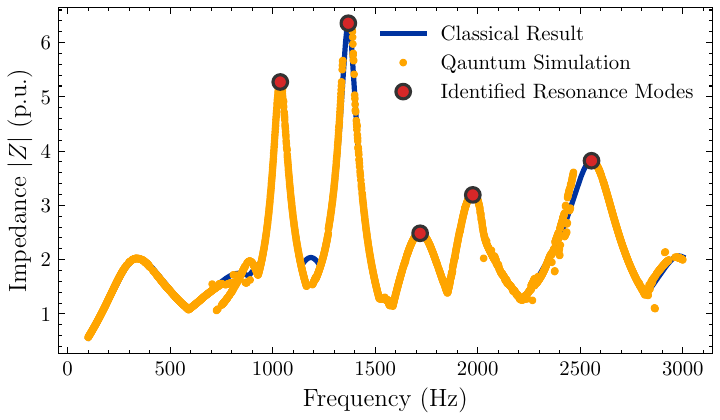}
    \caption{Validation of the quantum algorithmic framework on the 5-bus transmission system. The critical resonance modal impedance ($|Z_{cm}|$) is plotted across a continuous frequency spectrum. The solid blue line represents the exact classical solution derived via standard matrix diagonalization, while the red markers indicate the eigenvalues computed using the Quantum RVVQE algorithm. The near-perfect overlap confirms the accuracy of the quantum approach in capturing the singularities of the non-Hermitian admittance matrix.}
    \label{fig:resonance_comparison}
\end{figure}

To clearly outline this complete grid-to-quantum workflow, the step-by-step execution across the frequency spectrum is formalized in Algorithm \ref{alg:qrma}.

\begin{algorithm}
\caption{Quantum-Classical Hybrid RMA (Q-RMA) via RVVQE}
\label{alg:qrma}
\begin{algorithmic}[1]
\State \textbf{Input:} $n$-bus grid data, frequency range $[f^{(0)}, f^{(mx)}]$, step $\Delta f$, tolerance $\epsilon$
\State \textbf{Output:} Critical resonance modal impedances $Z_{cm,f}$
\State $f \leftarrow f^{(0)}$
\While{$f \le f^{(mx)}$}
    \State Compute non-Hermitian admittance matrix $Y_{B,f}$ from grid component models
    \State Decompose $Y_{B,f} = H_f + iK_f$, where $H_f, K_f$ are Hermitian
    \State Map $H_f, K_f$ to $N$-qubit Pauli strings ($N = \lceil \log_2 n \rceil$)
    
    \State Initialize parametrized quantum state $|\psi(\theta)\rangle$
    \Repeat
        \State Measure $\langle H_f\rangle, \langle K_f\rangle, \langle H_f^2\rangle, \langle K_f^2\rangle$ on the QPU
        \State Compute: $C_{var}(\theta)=\langle H_f^2\rangle_{\theta}-\langle H_f\rangle_{\theta}^2+\langle K_f^2\rangle_{\theta}-\langle K_f\rangle_{\theta}^2$
        \State $\theta \leftarrow$ Update parameters via classical optimizer
    \Until{$C_{var}(\theta) < \epsilon$}
    \State Optimal parameters $\theta^* \leftarrow \theta$
    \State Compute eigenvalue: $\lambda_{n,f} = \langle\psi(\theta^*)|H_f|\psi(\theta^*)\rangle + i\langle\psi(\theta^*)|K_f|\psi(\theta^*)\rangle$
    \State Record critical modal impedance: $Z_{cm,f} = 1 / \lambda_{n,f}$
    
    \State $f \leftarrow f + \Delta f$
\EndWhile
\State \textbf{return} Resonance profile $Z_{cm,f}$ array
\end{algorithmic}
\end{algorithm}

By using this structured method, we successfully extract the critical resonance frequencies spanning the entire grid spectrum using Hermitian measurements only, rendering the algorithm highly scalable on near-term hardware.


\section{Results and Discussion}

To validate the proposed quantum method, we used a standard 5-bus transmission system as a benchmark. Following the classical approach reported in earlier resonance-mode studies \cite{cartiel2023faster}, the non-Hermitian admittance matrices ($Y_{bus}$) were generated over a frequency range from 100 Hz to 3000 Hz. These matrices were obtained by combining analytical white-box models of grid components with measured or frequency-dependent black-box impedance/admittance models for components without analytical descriptions \cite{orellana2021stability}. Since the original admittance matrix is of size $5 \times 5$, it was padded with zeros to form an $8 \times 8$ matrix. This allows the matrix dimension to match $2^n$, making it suitable for representation on a 3-qubit quantum system. Before applying the quantum algorithm, the exact eigenvalues were calculated using the SciPy linear algebra library. These classical results were used as the reference values for the critical resonance modes ($|Z_{cm}|$).

The padded non-Hermitian matrices were then solved using the RVVQE algorithm \cite{pandey2026rvvqe}. First, each matrix was separated into two Hermitian matrices,
\begin{equation}
H=\frac{Y_{bus}+Y_{bus}^{\dagger}}{2}, \qquad
K=\frac{Y_{bus}-Y_{bus}^{\dagger}}{2i},
\end{equation}
which were further expressed as sums of tensor-product Pauli operators. A hardware-efficient parameterized quantum circuit (ansatz), shown in Fig.~\ref{fig:ansatz_3q_ent}, was used to prepare the trial quantum state $|\psi(\vec{\theta})\rangle$. To obtain all the eigenvalues, a sequential deflation technique was included in the cost function. The cost function minimizes the real variance while adding a penalty based on the overlap between the current trial state and the eigenstates that have already been found. This penalty ensures that each new solution is orthogonal to the previous ones. The circuit parameters were optimized using a gradient-based classical optimizer, and the complete workflow is summarized in Algorithm~\ref{alg:qrma}.

The optimization starts with an initial parameter vector $\vec{\theta}$ obtained from a grid search over the interval $[0,2\pi]$. This provides a better starting point for the optimizer and reduces the possibility of converging to local minima. During the calculation, an automatic validation routine removes zero eigenvalues and repeated degenerate solutions. When closely spaced eigenvalues are detected, a small diagonal shift is introduced to separate them. The orthogonality condition between previously obtained eigenstates is enforced through the cost function, allowing the algorithm to converge to different eigenstates without the convergence issues commonly observed in methods such as the Power Iteration algorithm.

Figure~\ref{fig:resonance_comparison} shows the resonance modal impedance ($|Z_{cm}|$) obtained using the proposed quantum approach together with the classical results. The two results are in close agreement over the entire frequency range, indicating that the RVVQE algorithm accurately reproduces the complex eigenvalues of the non-Hermitian admittance matrix.

The calculations presented in this work were performed without applying Quantum Error Mitigation (QEM). Even so, the quantum results remain close to the classical reference values. Since present-day quantum hardware is affected by gate errors and decoherence, the use of Quantum Error Mitigation techniques, such as Zero-Noise Extrapolation (ZNE) and Probabilistic Error Cancellation (PEC), is expected to further improve the accuracy of the computed eigenvalues.

The present work considers a 5-bus transmission system as a proof of concept. The same procedure can be applied to larger benchmark systems. As the size of the network increases, classical eigenvalue calculations become more demanding because of the increasing computational cost of large matrix operations \cite{saad2011numerical}. Future studies will investigate larger IEEE test systems and explore the use of Quantum Random Access Memory (QRAM) for efficiently representing large sparse admittance matrices on quantum hardware.
\section{Conclusion}

In this work, we presented a quantum-classical hybrid approach for Resonance Mode Analysis (RMA) of multi-terminal transmission grids. Instead of using only classical eigenvalue decomposition techniques, the proposed method employs the Real Variance-based Variational Quantum Eigensolver (RVVQE) to determine the resonance modes of the system \cite{pandey2026rvvqe}. The results obtained for the 5-bus transmission system show that the quantum algorithm accurately reproduces the critical resonance modes and closely matches the results obtained using classical methods.

One of the main advantages of the proposed approach is its potential for handling large power systems. In classical RMA, storing and diagonalizing large non-Hermitian admittance matrices requires significant computational memory and processing time. In contrast, quantum computing represents the system using a logarithmic number of qubits, reducing the memory required to represent large matrices. Although current quantum hardware is still limited, this property makes quantum algorithms attractive for future large-scale power system analysis.

The proposed framework can be extended to larger benchmark systems, such as the IEEE 118-bus and IEEE 300-bus test networks \cite{ieee118bus,ieee300bus}. With future improvements in quantum hardware and quantum error mitigation techniques, the same approach may also be applied to very large synthetic power grids, such as the ACTIVSg70k system \cite{activsg70k}. Overall, this work demonstrates that RVVQE is a promising quantum algorithm for solving the non-Hermitian eigenvalue problems that arise in resonance mode analysis and provides a foundation for future quantum applications in power system stability studies.
\section{acknowledgements}
This work is supported by the SERB-DST, Govt.~of India, via project \sloppy{CRG/2022/009359}.
We acknowledge the National Supercomputing Mission (NSM) for providing computing resources of `PARAM Ganga' at IIT Roorkee, which is implemented by C-DAC and supported by MeitY and DST, Govt.~of India.


\end{document}